\documentstyle[12pt]{article} \textwidth    165mm    \textheight 200mm
\begin{document}

\vspace*{0.5cm}
\centerline{PACS no.s 04.60.+n, 03.65.-w, 11.10.-z}
\centerline{Preprint IBR-TH-97-S-033, January 9, 1997}

\vspace*{1cm}
\begin{center}
{\large \bf ISOMINKOWSKIAN UNIFICATION OF THE \\
SPECIAL AND GENERAL RELATIVITIES}
\end{center}

\vskip 0.5cm
\centerline{{\bf Ruggero Maria Santilli}}
\centerline{Institute for Basic Research}

\centerline{P.O.Box 1577, Palm Harbor, FL 34682, U.S.A.}
\centerline{E-address ibr@gte.net; Web Site http://home1.gte.net/ibr}

\begin{abstract}
We submit  a   classical   unification of  the special    and  general
relativities via the new   isominkowskian  geometry in which  the  two
relativities are  differentiated by the basic unit.   We then show that
the unification admits an operator image in which gravitation verifies
the  abstract   axioms  of  relativistic  quantum mechanics    under a
universal symmetry   which  is    isomorphic to     the   Poincar\'{e}
symmetry.     The  compliance  of    the unification    with available
experimental data is indicated. This study has been permitted by the
recent achievement of sufficient mathematical
maturity in memoir$^{3f}$, axiomatic consistency in memoir$^{3t}$ and
generalized symmetry principles in memoir$^{4v}$. More detailed
studies are presented in the forthcoming paper$^{3u}$.
\end{abstract}

\vskip 0.5cm

{\bf 1. Introduction}. One of the most majestic
achievements of this century for mathematical
beauty, axiomatic  consistency and experimental verifications has been
the  {\it  special theory of  relativity}  (STR)$^{1}$. By comparison,
despite equally outstanding  achievements, the {\it  general theory of
relativity} (GTR)$^{2}$ has remained afflicted by numerous problematic
aspects at both classical and quantum levels.

In view of the above, in this note we submit a formulation of the {\it
general} relativity  via the axioms  of  the {\it  special}, under the
condition  of   preserving  Einstein's    field equations   and  their
experimental
verifications. Our study   is conducted via axiom--preserving maps   of
conventional   structures,   called  {\it   isotopies},   initiated in
ref.s$^{3}$  and studied  by various  authors$^{4,5}$.

 This study has been permitted by the
recent achievement of sufficient mathematical
maturity in memoir$^{3f}$, axiomatic consistency in memoir$^{3t}$ and
maturity in
generalized symmetry principles in memoir$^{4v}$. More detailed
studies are presented in the forthcoming paper$^{3u}$.

The property  at the foundation of  this  note is that  the transition
from the  Minkowskian metric $\eta = Diag (1, 1, 1, -1)$
to a (3+1)-dimensional  Riemannian metric g(x) is characterized  by a
{\it  noncanonical}  transformation  $x\rightarrow   x' =  U\times  x,
U\times  U^{t}\not =  I$, for  which (by ignoring  hereon the dash)
$g(x) = U\times\eta\times U^{t}$.

The above mathematically trivial occurrence has rather serious physical
implications
at both classical and quantum levels. In fact, at the classical level it
implies
that gravitational theories constructed over a Riemannian spaces do not
possess
invariant units of space and time (evidently because they are not
preserved by the noncanonical time evolution of the theory by
definition),
thus implying evident ambiguities in the application of the theory to
actual measurements (e.g., because it is not possible to conduct
releable measures, say, of length, with a stationary meter varying in
time)
and other problematic aspects$^{3t}$.

More generally, the noncanonical character of Riemannian theories of
gravitation implies that the still unresolved problematic
aspects debated during this century on gravitation, quite likely,
are not due to Einstein's field equations, but rather to the lack of
proper
mathematics used in their treatment, which is the viewpoint adopted
in this note.

At the operator level it is easy to see that, for consistency with the
classical
counterpart, quantum gravity must have a nonunitary time evolution when
referred
to a conventional Hilbert space over a conventional field. It is now
established$^{3t}$ that theories with nonunitary time evolutions are
afflicted by the following shortcomings: 1) They do not have invariant
units
of space and time, by therefore lacking physically un-ambiguous
applications to
experimental measurements; 2) they do not preserve Hermiticity in time,
thus lacking physically acceptable observabnles; and 3) they
do not have unique and
invariant numerical predictions (because of the lack of uniqueness and
invariance of the special functions needed for data elaboration).

In an attempt to resolve the above problematic aspects in due time, in
this
note we submit a novel formulation of gravity based on the following
central assumptions: 1) Preservation unchanged of Einstein's field
equations
and related experimental verifications; 2) classical formulation of
these
equations under the uncompromisable conditions of possessing invariant
units of space and time, as it is the case for the special relativity;
and 3)
operator formulation of gravity based on the axioms of conventional
relativistic quantum mechanics.

\vskip 0.5cm

{\bf 2. Isominkowskian geometry}. A study of the above conditions
is permitted by a novel form of mathematics called {\it isomathematics}
originally
proposed by Santilli$^{3a}$ and then studies in various works$^{3-5}$.
It is
characterized by liftings of conventional
mathematics called {\it isotopies} which map any linear local
and canonical structure
into its most general possible nonlinear, nonlocal and noncanonical
form,
which is nevertheless capable of reconstructing linearity, locality and
canonicity
on certain generalized spaces and fields called
{\it isospaces} and {\it isofields}.

The isotopies are ideally suited to study assumptions 1), 2) and 3) of
Sect. 1
because they are axiom-preserving by conception and construction, while
the
 new formulations are locally isomorphic to the original ones. This
evidently
ensures the preservation unchanged of Einstein's axiom {\it ab initio}.
Jointly, the isotopies do permit the achievement of a theory with a
basic
invariant unit, as we shall see.

The fundamental isotopy for relativistic theories is the lifting  of the
unit of conventional theories, the unit I = diag. (1, 1, 1, 1) of the
Minkowski space and of the Poincare' symmetry,
into a  well  behaved, nowhere singular,
Hermitean and   positive--definite  $4 \times 4$-dimensional
matrix $\hat {I}$    whose elements   have  an
arbitrary dependence on local quantities and, therefore, can depend on
the  space--time coordinates x and other needed variables,
$I\rightarrow \hat{I}  =  \hat{I}(x)>0$,
while the conventional  associative product $A\times B$  among generic
quantities  A,  B is lifted   by  the {\it  inverse}  amount, $A\times
B\rightarrow A\hat{\times}  B  = A\times  \hat{T}\times B,  \hat{I}  =
\hat{T} ^{-1}$.

Under these   assumptions $\hat{I}$ is the  (left and
right) generalized unit of  the  new theory, $\hat{I}\hat{\times}  A =
A\hat{\times}\hat{I} \equiv A,  \forall A$,  called the {\it  isounit}
and $\hat{T}$  is called the  {\it isotopic element}. For consistency,
the {\it totality} of the  original theory must then be  reconstructed
to admit $\hat{I}$ as the correct (left and  right) unit. This implies
the  isotopies of    numbers,  angles,  fields,  spaces,  differential
calculus,  functional   analysis, geometries,   algebras,  symmetries,
etc.$^{3r}$ (see ref.$^{3f}$ for a recent account).

Let $M(x, \eta, R)$ be the  Minkowski space with  space--time
coordinates
$x = \{x^{\mu}\} = \{r, x^{4}\}$, $x^{4} = c_{0}t$ (where $c_{0}$
is the speed of
light in vacuum), and  metric $\eta =  Diag. (1,1,1,-1)$ over the reals
$R  =  R(n,+,\times)$. Let   $\Re(x,g(x),R)$ be  a $(3+1)$--dimensional
Riemannian   space with nowhere singular   and  symmetric metric $g  =
g^{t} = U\times\eta\times U^{t}$.

A study of conditions 1), 2), and 3) of Sect. 1 is then permitted
by assuming as basic isounit
of  our theory the quantity  $\hat{I} = U\times U^{t} = \hat{I}^{t}>0$
with explicit form derivable form a Riemannian metric via the {\it
isominkowskian factorization}$^{3o,3p}$

\vskip 0.5cm
\begin{equation}
g(x) =   \hat{T}(x)\times\eta,  \hat{I}(x) = [T(x)]^{-1}   = U\times
U^{t}.
\label{eq:one}\end{equation}

\vskip 0.5cm
As  an example, for    the  case  of the  celebrated   Schwarzschild's
metric$^{2d}$,  we   have $U\times U^{t}  =  \hat{I}  =  Diag. ((1-M/r),
(1-M/r), (1-M/r), (1-M/r)^{-1})$ and similarly for other metrics. Note
that  the positive--definiteness of $\hat{I}$  is assured  by the
locally
Minkowskian   character of Riemann. Without   loss  of generality, the
isounit can therefore be assumed herein as being diagonal.

To construct    the  {\it isotopies of the    STR},  also called  {\it
isospecial relativity}$^{3h-3t}$, we first  need the lifting of numbers
and  fields$^{3g}$. For   this    we  note  that   the    conventional
multiplicative   unit I is lifted   into  the isounit, $I  \rightarrow
U\times 1\times U^{t}    = \hat{I}$ while   the  additive unit  $0$
remains   unchanged,   $0  \rightarrow   \hat{0}  =  U\times  0\times
U^{t} =  0$.  The  numbers are  lifted   into the  so--called {\it
isonumbers},  $n \rightarrow  \hat{n}   =   U\times n\times U^{t}    =
n\times\hat{I} = (n \times m)\times \hat{I}$
with lifting of the product $n\times m
\rightarrow \hat{n}\hat{\times}\hat{m} = \hat{n}\times\hat{T}\times
\hat{m}$.

The original  field $R=R(n,+,\times  )$ is  then lifted  into the
isofield
$\hat{R}=\hat{R}  (\hat{n},\hat{+}.\hat{\times   })$   for  which  all
operations  are isotopic It  is easy to  see that $\hat{R}$ is locally
isomorphic  to   $R$ by   construction   and,  thus,  the  lifting  $R
\rightarrow  \hat{R}$ is   an  isotopy. Despite   its  simplicity, the
lifting is not  trivial, e.g., because the  notion of primes and other
properties of  number theory depend on the  assumed unit.  For further
aspects we refer to$^{5r}$ which also includes the isotopies of angles
and    functions  analysis.  Note  for   later   needs the  identity,
$\hat{n}\hat{\times }A\equiv n\times A$.

Next, we  need  the   lifting  of    the space  $M$   into  the   {\it
isominkowskian     space}  $\hat{M}   =      \hat{M}(\hat{x},\hat{\eta
},\hat{R}$)  first proposed by Santilli in Ref.$^{3h}$
which is characterized by the {\it isocoordinates} $x
\rightarrow \hat{x} =  U\times  x\times U^{t} = x\times  \hat{I}$, and
{\it  isometric} $\eta\rightarrow \hat{\eta  }(x) = U\times \eta\times
U^{t} \equiv  g(x)$ although,   for  consistency, the latter   must be
defined on $\hat{R}$, thus having the structure $\hat{N} =
(\hat{N}_{\mu\nu}) =
\hat{\eta}\times\hat{I}=(\hat{\eta}_{\mu\nu})\times\hat{I}$.      The
conventional  interval  on  $M$ is then   lifted  into the  {\it
isointerval} on $\hat{M}$ over $\hat{R}^{4h}$

\vskip 0.5cm
\begin{eqnarray}
\lefteqn{(\hat{x}-\hat{y})^{\hat{2}}=(\hat{x}-\hat{y})^{\mu}
\hat{\times}\hat{N}_{\mu\nu}\hat{\times}(\hat{x}-\hat{y})^{\nu} =
 [(x-y)^{\mu}\times\hat{\eta}_{\mu\nu}\times
(x-y)^{\nu}]\times\hat{I}  =} \nonumber \\[0.5cm]
&=& [(x^{1}-y^{1})\times\hat{T}_{11}\times(x^{1}-y^{1})+
(x^{2}-y^{2})\times T_{22}\times(x^{2}-y^{2})+ \nonumber \\[0.5cm]
&&+(x^{3}-y^{3})\times T_{33}\times(x^{3}-y^{3})-(x^{4}-y^{4})
\times T_{44}\times(x^{4}-y^{4})\times\hat{I}.
\label{eq:two}\end{eqnarray}
\vskip 0.5cm

As one can see, the above interval coincides with the conventional
Riemannian interval byu conception, except for the factor $\hat{I}$.

It  easy to see that $\hat{M}$  is locally isomorphic  to  $M$ and the
lifting $M \rightarrow \hat{M}$ is also an isotopy. In particular, the
isospace  $\hat{M}$ is {\it isoflat},  i.e.,  it verifies the axiom of
flatness {\it in isospace over the isofields}, that is, when referred
to the generalized  unit  $\hat{I}$, otherwise $\hat{M}$ is  evidently
curved owing  to the  dependence $\hat{\eta}=\hat{\eta}(x) = g(x)$.  In
other
words,  {\it assumptions (1)  eliminate the curvature while preserving
the Riemannian metric}. As we shall  see, this appears to be essential
to achieve a gravitational  theory with an  invariant basic unit. Note
that   $\hat{M}$ and   $\hat{R}$    have   the  {\it same     isounit}
$\hat{I}$. Studies of {\it isocontinuty  properties} on isospaces have
been  conducted by Kadeisvili$^{4r}$  and  those   of
the  underlying novel  {\it isotopology} by Tsagas and Sourlas$^{4s}$.

The {\it isominkowskian geometry}$^{3r}$ is  the geometry of isospaces
$\hat{M}$,  and incorporates in a   symbiotic way both the Minkowskian
and   Riemannian geometries.   In   fact, it  preserves all  geometric
properties of the  conventional {\it Minkowskian}  geometry, including
the light cone  and the  maximal causal speed $c_{o}$ (see below),
while jointly incorporating  the machinery of the Riemannian geometry
in an isotopic form. As such, it is ideally suited for our objectives.

It should be indicated that this author has studied until
the {\it interior gravitational problem} via the {\it isotopies of the
Riemannian geometry}. The use of the {\it isominkowskian} geometry for
the characterization of {\it the exterior gravitational problem}
was brifly indicated in note$^{3o}$
and it is studied in more details in this work. Also, the main line of
the
isominkowskian geometry inclusive of the machinery of the Riemannian
geometry are presented in this study for the first time.

To outline the new geometry, one must know that, unexpectedly, the use
of  the  ordinary differential  calculus leads  to  inconsistencies
under
isotopies (e.g., lack    of   invariance) because  dependent    on the
assumption of the trivial unit 1 in a hidden way.  The central tool of
the isominkowskian   geometry  is therefore  the  {\it isodifferential
calculus} on $\hat{M}(\hat{x},\hat{\nu},\hat{R})$,   first  introduced
in$^{3g}$, which   is   characterized by  the  {\it  isodifferentials,
isoderivatives} and related properties $\hat{d}x^{\mu } = \hat{I}^{\mu
}_{\nu }\times dx^{\nu},  \hat{d}x_{\mu } = \hat{T}_{\mu }^{\nu
}\times dx_{\nu},
\hat{\partial }_{\mu }=\hat{\partial }/\hat{\partial }x^{\mu} =
\hat{T}_{\mu }^{\nu }
\times\partial /\partial x^{\nu },
\hat{\partial }^{\mu } = \hat{\partial }/\hat{\partial }x_{\mu } =
\hat{I}^{\mu }_{\nu }\times\partial /\partial x_{\nu },
\hat{\partial }x^{\mu }/\partial x^{\nu } = \delta^{\mu }_{\nu },
\hat{\partial }x_{\mu }/\hat{\partial }x^{\nu } = \hat{\eta }_
{\mu\alpha }\hat{\partial }x^{\alpha }/\hat{\partial }x^{\nu } =
\hat{\eta }_{\mu\nu },
\hat{\partial }^{\mu }/\hat{\partial }x_{\nu } =
 \hat{\eta }^{\mu\alpha }\times\hat{\partial }x_{\alpha }/
\hat{\partial }x^{\nu } = \hat{\eta }^{\mu\nu }$.

Note that the original axioms must be preserved  for an isotopy. Thus,
$\hat{\partial }_{\alpha  }\hat{\partial}\_{\beta  } = \hat{\partial
}_{\beta  }\hat{\partial }_{\alpha  }$ and,  therefore, $\hat{\partial
}_{\alpha    }\hat{\partial    }_{\beta   }    =    T_{\alpha   }^{\mu
}\times\hat{T}_{\beta }^{\nu }\times\partial _{\mu }\partial _{\nu }$.

Note    also  the    hidden  {\it   isoquotient}$^{3g}   A/\hat{}B   =
(A/B)\times\hat{I}$ and isoproduct  $\hat{\partial}
\hat{\times} \hat{\partial}$. Thus,
by including the isoquotient, the quantity $\hat{\partial
}\hat{\partial}$
should be moe rigorously
written $\hat{\partial }\hat {\times} \hat{\partial}$. This results in
an
inessential final multiplication of the expression considered
-by $\hat {I}$ and,
as such, it will be ignored hereon for simplicity.

The entire formalism   of the {\it  Riemannian} geometry  can  then be
formulated on the  {\it iso\-min\-kow\-ski\-an} space via the
isodifferential
calculus. This aspect is  studied in details elsewhere$^{3t}$. We here
mention: {\it isochristoffel's symbols} $\hat {\Gamma}
_{\alpha\beta\gamma} =
\hat{\frac{1}{2}}\hat\times  (\hat{\partial}   _{\alpha    }\hat{\eta
}_{\beta\gamma } +
\hat{\partial }_{\gamma }\hat{\eta }_{\alpha\beta } -
\hat{\partial }_{\beta }\hat{\eta }_{\alpha\gamma })\times \hat{I}$,
{\it isocovariant differential}
$\hat{D}\hat{X}^{\beta} = \hat{d}\hat{X}^{\beta} +
\hat{\Gamma }_{\alpha }^{\beta }{}_{\gamma }\hat {\times}
\hat{X}^{\alpha }\hat {\times} \hat{d}\hat{x}^{\gamma }$,
{\it isocovariant       derivative}
 $\hat{X}^{\beta}_{|\hat{}\mu} =
\hat{\partial }_{\mu }\hat{X}^{\beta} +
\hat{\Gamma }_{\alpha }^{\beta }{}_{\mu } \hat {\times} \hat{X}^{\alpha
}$,
{\it isocurvature tensor}
$\hat{R}_{\alpha }^{\beta }{}_{\gamma\delta }
=\hat{\partial }_{\beta }\hat{\Gamma }_{\alpha }^{\beta }{}_{\gamma } -
\hat{\partial }_{\gamma }\hat{\Gamma }_{\alpha }^{\beta }{}_{\delta } +
\hat{\Gamma }_{p}^{\beta }{}_{\delta } \hat {\times} \hat{\Gamma
}_{\alpha }^
{p}{}_{\gamma } -
\hat{\Gamma }_{p}^{\beta }{}_{\gamma } \hat {\times} \hat{\Gamma
}_{\alpha }^
{p}{}_{\delta }$,
etc.

The verification, this time,  of the {\it Riemannian} properties
is shown by  the  fact that (under   the assumed conditions)  {\it the
isocovariant derivatives  of all isometrics  on } $\hat{M}$ {\it over}
$\hat{R}$ {\it   are   identically null},  $\hat{\eta   }_{\alpha\beta
|\hat{}\gamma} \equiv 0,\alpha , \beta , \gamma  = 1, 2,  3, 4$.  This
illustrates that    the   Ricci Lemma  also    holds  under the   {\it
Minkovskian} axioms.

A  similar occurrence holds  for all other axioms
of the  Riemannian  geometry  (including   the forgotten   {\it  Freud
identity}, as we shall study in more detail elsewhere$^{3u}$.

\vskip 0.5cm

{\bf 3. Clssical unification of the special and general relativities}.
We
are now equipped to present, apparently for the first time, the
classical equations of our isominkowskian formulation
of gravity, here called
{\it  isoeinstein equations} on $\hat{M} over \hat{R}$,
which can then be written

\vskip 0.5cm
\begin{equation}
\hat{G}_{\mu\nu } = \hat{R}_{\mu\nu } -
\hat{\frac{1}{2} }\hat{\times }\hat{N}_{\mu\nu }\times\hat{R} =
\hat{k} \hat{\times } \hat{\tau }_{\mu\nu },
\label{eq:three}\end{equation}

\vskip 0.5cm
where $\hat{\tau  }_{\mu\nu }$   is  the  source {\it   isotensor}  on
$\hat{M},\hat{\frac{1}{2} } = \frac{1}{2}\times\hat{I},
\hat{N}_{\mu\nu } = \hat{\eta }_{\mu\nu }\times\hat{I} =
g_{\mu\nu }\times\hat{I},   \hat{k}  = k\times \hat{I}$  and  $k$
is the  usual constant.

Despite apparent differences, it should be  indicated that
Eq.s~(3)
{\sl  coincide numerically with  Einstein's  equations} both in
isospace as well as in their projection in ordinary spaces for all
diagonal Riemannian metrics. In fact, in which case $\hat{T}$ is
also diagonal with $\hat{\eta}\equiv g(x)$.

In isospace, the isoderivative $\hat{\partial}_\mu  = \hat{T}_\mu^\alpha
\times\partial_\alpha$   deviates  from   the  conventional derivative
$\partial_\mu$ by the isotopic factor $\hat{T}$.
But its numerical value  must  be referred to $\hat{I} =
\hat{T}^{-1}$, rather than $I$. This implies the preservation in
isospace of
the original value of$\partial_\mu$ and, consequently, of the original
field
equations.

For the case of the projection of in ordinary spaces, the isoquations
are
reducible to the conventional equations multiplied by common isotopic
factors which, as such, are inessential and can be eliminated. In fact,
the isochristoffel's symbols~ deviate  from  the
conventional symbols by the same factor $\hat{T}$ (again, because
$\hat{\eta}
\equiv    g$), and the same happens with other terms, except for
possible redefinition of the source when needed, thus preserving again
the
conventional field equations and related experimental verifications.

A  more detailed  study  of  Eq.s~(3)  and  related  isominkowskian
geometry is  presented in  the forthcoming Ref.$^{3u}$,
including  the use  of the
forgotten {\sl Freud identity} of  the Riemannian geometry in its
isominkowskian realization.

In  summary, the isominkowskian formulation  of gravity permits a {\it
geometric unification  of the special  and general relativity into one
single relativity, the isospecial relativity}$^{3s}$ where for $\hat{I}
= I =    Diag.(1,1,1,1)$  we have   the  special and   for $\hat{I}  =
\hat{I}(x) = U\times U^{t}$ we have the general. The invariance of the
isounit is  illustrated below. More detailed  studies are available in
the forthcoming paper$^{3t}$.

\vskip 0.5cm

{\bf 4. Operator unification of the special and general relativities.}
We
now  indicate that the above   {\it classical} unification admit a
step--by--step {\it operator}   counterpart, here called   {\it operator
isogravity} (OIG). It should be indicated from the  outset that OIG is
structurally different than  the  conventional {\it quantum   gravity}
(QG)$^{6}$ on   numerous grounds, e.g., because  OIG  and QM have {\it
different  units, Hilbert}   spaces,   etc. In particular,  the   word
"operator" in OIG  is suggested to keep in  mind  the differences with
"quantum" mechanics (as it should also be for QG).

To identify  OIG,   we  note that   the  original  {\it  noncanonical}
transform $U\times  U^{t}  = \hat{I}\not{=}  I $ is  mapped into a  {\it
nonunitary} transform  on a conventional Hilbert  space $\cal  H$ over
the complex field $C(c,+,\times)$.  The isounit of the operator theory
is therefore $\hat{I} = U\times U^{\dagger} =  \hat{I}, \hat{T}
= (U\times U^{\dagger })^{-1} = T^{\dagger } = \hat{I}^{-1}$, where the
representation of gravity occurs  as per Eq.s  (1). Then, OIG requires
the  isotopies of the  {\it  totality}  of  {\it relativistic  quantum
mechanics} (RQM) resulting in a formulation known as {\it relativistic
hadronic mechanics} (RHM)$^{3t}$.

Besides   the  preceding   isotopies    $R\rightarrow  \hat{R}$    and
$\hat{M}\rightarrow\hat{M}$, RHM is based on the lifting of the Hilbert
space $\cal H$  with states  $|\Psi >,|\Phi >,  ...$ and inner  product
$<\Phi |\Psi   >\in C(c,+,\times )$  into the   {\it isohilbert space}
$\hat{\cal H}^{4t}$ with  {\it isostetes} $|\hat{\Psi }> = U\times|\Psi
>, |\hat{\Phi  }> =  U\times  |\Phi >,   ...,$ {\it isoinner  product}
$<\hat{\Phi }|\hat{}\hat{\Psi  }>   = U\times   <\Phi |\Psi    >\times
U^{\dagger   }    =  <\hat{\Phi   }|\times\hat{T}\times  |   \hat{\Psi
}>\times\hat{I}$       defined       on         the           isofield
$\hat{C}(\hat{c},\hat{+},\hat{\times     })$   with   isonormalization
$<\hat{\Psi }|\times\hat{T}\times |  \hat{\Psi }> =  1$.

 We then have
the  {\it iso--four--momentum    operator}$^{3s}$ $p_{\mu   }\hat{\times
}|\hat{\Psi }>  =    -\hat {i} \hat {\partial}_{\mu}|\hat{\Psi }>$,
with   {\it
fundamental isocommutation rules} $[\hat{x}_{\mu }, \hat{}\hat{p}_{\nu
}] = U\times [x_{\mu }, p_{\mu }]\times U^{\dagger} =
\hat{x}_{\mu }\times\hat{T}\times\hat{p}_{\nu } -
\hat{p}_{\nu }\times\hat{T}\times\hat{x}_{\mu} =
\hat{i}\hat{\times }\hat{N}_{\mu\nu }$.
The  (nonrelativistic) {\it  isoheisenberg' equations}$^{3b}$ and {\it
isoschrodinger equations}$^{3t,3u}$ can the  be  written, in terms  of
the isodifferential calculus of ref.$^{3g}$

\vskip 0.5cm
\begin{eqnarray}
&&\hat{i}\hat{\times}\hat{d}A/\hat{d}t = i\times\hat{I}_{t}\times dA/dt
= [A,\hat{}H] =  A\times\hat{T}_{s}\times H - H\times\hat{T}_{s}\times
A, \;
\hat{I} = \hat{I}_{s}\times\tilde{I}_{t}, \nonumber \\[0.5cm]
&&\hat{i}\hat{\times}\hat{\partial}_{t}|\hat{\Psi }> =
i\hat{I}_{t}\times\partial_{t}|\hat{\Psi}> =  H\times_{s}|\hat{\Psi }>
=         \nonumber  \\[0.5cm]
& &\;\;\; = H\times\hat{T}_{s}\times|\hat{Psi          }>             =
\hat{E}\hat{\times_{s}}|\hat{\Psi}>                                  =
E\times\hat{I}_{s}\times\hat{T}_{s}\times|\hat{\Psi } > \equiv E\times
|\hat{\Psi }>.
\label{eq:four}\end{eqnarray}

\vskip 0.5cm
Note that the  final numbers of the theory  are conventional. We  also
have the lsifting of  expectation values into the form $\hat{<}A\hat{>}
= <\hat{\Psi }|\times
\hat{T}\times A\times\hat{T}\times |\hat{\Psi } >
/< \hat{\Psi }|\times\hat{T}\times  |\hat{\Psi }>$, and the  compatible
liftings  of the  remaining   aspects of RQM$^{3t}$.  In   particular,
$\hat{I}$     is  the  fundamental    invariant     of    the
isotheory,
$i\hat{d}\hat{I}/\hat{d}t     =  \hat{I}\hat{\times    }H     -
H\hat{\times}\hat{I}\equiv 0$.

It  is easy   to prove that   RHM  preserves  {\it  all}  conventional
properties of RQM$^{4t}$. In particular: {\it isohermiticity coincides
with conventional hermiticity},  $H^{\dagger} \equiv H^{\dagger}$  (all
quantities which  are originally   observables remain,  therefore,  so
under  isotopies); {\it the  isoeigenvalues  of isohermitean operators
are isoreal}   (thus   conventional because of    the    identity
$\hat{E}\hat{\times}|\hat{\Psi }> \equiv E\times |\hat{\Psi }>)$; {\it
RHM is form invariant under isounitary transforms} $\hat{U}\hat{\times}
\hat {U}^{\dagger } = \hat {U}^{\dagger }\hat{\times }\hat{U}  =
\hat{I}$.
In fact,
we have the invariance of the isounit $\hat{I}\rightarrow \hat{I}'=
\hat{U}\hat{\times }\hat{I}\hat{\times }hat{U}^{\dagger }\equiv
\hat{I}$,
of the  isoassociative   product    $\hat{U}\hat{\times }(A\hat{\times
}B)\hat{\times}\hat{U}^{\dagger } =
\bar{A}\hat{\times }\bar{B}$;
etc;  and  the   same occurs for    all   other properties  (including
causality).  Note that nonunitary transforms on $\cal H$ can always be
identically rewritten as isounitary transforms on $\hat{\cal  H }, U =
\hat{U}
\times \hat{T}^{1/2}, U\times U^{\dagger } \equiv
\hat{U}\hat{\times }\hat{U}^{\dagger } =
\hat{U}^{\dagger }\hat{\times }\hat{U} = \hat{I}$, under which
RHM is invariant$^{3t}$.

It should be stressed that RHM {\it is not a new  theory, but merely a
new realization of the abstract  axioms of RQM}.  In fact, RHM and RQM
coincide at    the   abstract, realization--free    level   where   all
distinctions are  lost between $I$ and $\hat{I},  R$ and  $\hat{R}, M$
and $\hat{M},  \cal H$ and $\hat{\cal H}$, etc. Yet, RHM is broader than
RQM,  it  recovers the latter  identically  for $\hat{I}  = I$ and can
approximate the latter as close as desired for $\hat{I} \approx I$.

On summary, the entire formulation of RHM of memoir$^{3t}$ can be
consistently specialized for the gravitational isounit $\hat {I}(x)$
yielding the proposed OIG.

\vskip 0.5cm

{\bf 5. The universal isopoincar\'{e} symmetry of gravitation.} An
important property of the isominkowskian formulation of
gravity, which is lacking  for conventional formulations, is that of
{admitting a universal, classical and operator   symmetry for all
possible
Riemannian formulations of gravitation}  first
identified by Santilli$^{3h--3l}$ under the name of
{\it isopoincar\'{e}  symmetry}   $\hat{P}(3.1)$, {\it   which results
to  be
locally isomorphic to the conventional symmetry} $P(3.1)^{3h-3l}$.

The
isosymmetry can   be  easly constructed via   the   isotopies of Lie's
theory$^{3a,3d}$    called  {\it  Lie--Santilli   isoteory}$^{5}$ which
essentially  consists in the reconstruction of   all branches of Lie's
theory   (universal enveloping  algebras,  Lie   algebras,  Lie group,
transformation and  representation theories, etc.) for the generalized
unit $\hat{I}  = [\hat{T}(x)]^{-1}$. Since $\hat{I} >  0$, one can see
from the inception that the isopoincar\'{e} symmetry is isomorphic to
the
conventional one, $\hat{P}(3.1) \approx P(3.1)$ (see ref.$^{4t}$ for a
recent accounts).

The operator version of the isopoincar\'{e} symmetry
is  characterized  by the
conventional generators and parameter opnly lifted into isospaces over
isofields  $X   = \{X_{k}\}  = \{M_{\mu\nu}   =
x_{\mu}p_{\nu} - x_{\nu}p,p_{\alpha }\}
\rightarrow \hat{X} = \{\hat{M}_{\mu\nu} = \hat{x}_{\mu}\times
\hat{p}_{\nu} -
\hat{x}_{\nu}\times \hat{p}_{\mu},\hat{p}_{\alpha}\}, k = 1,2,...,10,
\mu ,\nu = 1,2,3,4,$
and $w = \{w_{k}\} = \{(\theta ,v),a\}  \in R \rightarrow \hat{w}
= w\times \hat{I} \in \hat{R}(\hat{n},+,\hat{\times })$. The isotopies
preserve the  original  connectivity properties$^{3r}$.  The connected
component   of  $P(3.1)$    is  then   given   by  $\hat{P}_{0}(3.1)   =
S\hat{O}(3.1)\hat{\times }\hat{T}(3.1)$, where $S\hat{O}(3.1)$ is  the
{\it  connected  isolorentz group}$^{3h}$  and  $\hat{T}(3.1)$ is  the
group   of {\it   isotranslations}$^{3k}$. $\hat{P}_{0}(3.1)$   can be
written via the {\it  isoexponentiation}  as $\hat{e}^{A} = \hat{I}  +
A/1! + A\hat{\times }A/2! + . . . = (e^{A\times\hat{T}})\times\hat{I}$
characterized  by   the      {\it   isotopic
Poincar\'{e}--Birkhoff--Witt
theorem}$^{3a,3d,5}$ of  the    underlying isoenveloping   associative
algebra

\vskip 0.5cm
\begin{equation}
\hat{P}_{0}(3.1):\hat{A}(\hat{w})  =
\Pi_{k}\hat{e}^{i\times X\times w}               =
(\Pi_{k}e^{i\times X \times\hat{T}\times w}) \times \hat{I} =
\tilde{A}(w)\times \hat{I}.
\label{eq:five}\end{equation}

\vskip 0.5cm
Note the  appearance of the  gravitational isotopic element $\hat{T}(x)$
in the {\it exponent} of   the group structure. This illustrates   the
nontriviality   of the lifting and its   {\it nonlinear} character, as
evidently   necessary  for   any symmetry  of    gravitation. What  is
intriguing  is that the   isopoincar\'{e} symmetry recovers linearity
on
$\hat{M}$ over $\hat{R}$, a property called  {\it isolinearity}$^{3t}$.

Conventional linear transforms on   $M$ {\it violate} isolinearity  on
$\hat{M}$  and  must then be    replaced with the {\it  isotransforms}
$\hat{x}' = \hat{A}(\hat{w})\hat{\times }\hat{X} =
\hat{A}(\hat{w})\times\hat{T}(x)\times\hat{x}$
which can be written   from  (5)  for computational purposes    (only)
$\hat{x}'  =   \tilde{A}(w)\times\hat{x}$.    The  preservation of   the
original      dimension   is   ensured     by     the  {\it   isotopic
Baker--Campbell--Hausdorff    Theorem}$^{3a,3d,5}$.  Structure  (5) then
forms a connected {\it Lie--Santilli isogroup}$^{5}$ with laws
$\hat{A}(\hat{w})\hat{\times}\hat{A}(\hat{w}') =
\hat{A}(\hat{w}')\hat{\times}\hat{A}(\hat{w}) = \hat{A}(\hat{w} +
\hat{w}'),\hat{A}(\hat{w})\hat{\times}\hat{A}(-\hat{w}) = \hat{A}(0) =
\hat{I}(x) = [T(x)]^{-1}$.

As one can see,  $\hat{P}_{0}(3.1)$ is {\it  noncanonical} on $M$ over
$R$ (e.g., because it {\it does not}  preserve the conventional unit I),
but it is canonical  on $\hat{M}$  over  $\hat{R}$, a  property called
{\it isocanonicity} (because  it leaves invariant by  construction the
isounit).  This confirms  the achievement,  apparently  for the first
time, of an
operator theory of gravity verifying the fundamental invariance of its
unit. The invariance at the classical level is consequential.

The  {\it     isodiscrete     transforms}$^{3i}$   are    given      by
$\hat{\pi}\hat{\times}x =
\pi\times x = (-r,x^{4}), \hat{\tau }\hat{\times }x = \tau\times x =
(r,-x^{4})$,
where $\hat{\pi } = \pi\times\hat{I},  \hat{\tau } = \tau\times\hat{I}$,
and  $\pi$, $\tau$ are the   conventional inversion operators. Despite
such a simplicity, the physical implications are nontrivial because of
{\it the possibility    of  reconstructing as  exact
discrete  symmetries when believed  to be  broken}, which is studied by
embedding all symmetry breaking
terms in the isounit$^{3s}$. One should be aware that this is a rather
general property of the Lie--Santilli  isotheory, thus holding also for
continuous  symmetries. In fact, contrary  to a  popular beliefs, this
note shows that  {\it the Lorentz  and Poincar\'{e} symmetries  are
exact
for gravitation}.

The  use    isodifferential calculus on   $\hat{M}$   then  yields the
Lie--Santilli isoalgebra $\hat{p}_{0}(3.1)^{3k}$

\vskip 0.5cm
\[
[\hat{M}_{\mu\nu     },\hat{}\hat{M}_{\alpha\beta  }]   =   i  \times
(\hat{\eta }_{\nu\alpha } \times \hat{M}_{\mu\beta } -
\hat{\eta }_{\mu\alpha } \times \hat{M}_{\nu\beta } -
\hat{\eta }_{\nu\beta } \times \hat{M}_{\mu\alpha } +
\hat{\eta }_{\mu\beta } \times \hat{M}_{\alpha\nu }),
\]
\begin{equation}
[\hat{M}_{\mu\nu }, \;\;   \hat{p}_{\alpha   }]  = i   \times
(\hat{\eta
}_{\mu\alpha } \times \hat{p}_{\nu } -
\hat{\eta }_{\nu\alpha} \times \hat{p}_{\mu}),
[\hat{p}_{\alpha },\hat{}\hat{p}_{\beta }] = 0, \;
\hat{\eta }_{\mu\nu } = g_{\mu\nu }(x),
\label{eq:six}\end{equation}

\vskip 0.5cm
where         $[A,\hat{}B]   =     A\times\hat{T}(x)\times     B     -
B\times\hat{T}(x)\times  A$   is   the  {\it   isoproduct} (originally
proposed in $^{3b}$),  which does  indeed satisfy  the Lie axioms   in
isospace, as one can  verify. Note the   appearance of the  Riemannian
metric $\hat{\eta  }_{\mu\nu }  =  g_{\mu\nu }(x)$,  this time, as  the
"structure      functions"   $\hat{\eta   }_{\mu\nu      }$   of    the
isoalgebra$^{3a,3d,5}$.  Note also  that the  {\it momentum components
isocommute} (while they are  notoriously non--commutative for QG). This
confirms the achievement of an isoflat representation of gravity.

The local isomorphism $\hat{p}_{0}(3.1) \approx p_{0}(3.1)$ is ensured
by the  positive--definiteness  of $\hat{T}$. In  fact,  the use of the
generators    in  the  form $\hat{M}^{\mu   }_{\nu    } = \hat{x}^{\mu
}\hat{\times }p_{\nu} -
\hat{x}^{\nu}\hat{\times}\hat{p}_{\mu }$ would yield the {\it
conventional}
structure constants under a {\it generalized} Lie  product, as one can
verify.  The  above   local isomorphism is   sufficient,  per se',  to
guarantee the axiomatic consistency of OIG.

The {\it isocasimir invariants}  of $\hat{p}_{0}(3.1)$ are  the simple
isotopic    image  of  the  conventional    ones  $C^{0}  =  \hat{I} =
[\hat{T}(x)]^{-1}, C^{(2)} = \hat{p}^{\hat{2}} =
\hat{p}_{\mu }\hat{\times }\hat{p}^{\mu } =
\hat{\eta }^{\mu\nu }\hat{p}_{\mu}\hat{\times}\hat{p}_{\nu}, C^{(4)} =
\hat{W}_{\mu }\hat{\times }\hat{W}^{\mu }, \hat{W}_{\mu } =
\in _{\mu\alpha\beta\pi }\hat{M}^{\alpha\beta }\hat{\times}\hat{p}^{\pi
}$.
>From     them, one can   construct     any needed {\it   gravitational
relativistic equation}, such as the {\it isodirac equation}

\vskip 0.5cm
\[
(\hat{\gamma     }^{\mu     }\hat{\times    }\hat{p}_{\mu     }    +
\hat {i} \hat {\times}\hat{m})\hat{\times
}|>  =   [\hat{\eta    }_{\mu\nu   }(x)  \times
{\hat{\gamma }}^{\mu}(x) \times \hat{T}(x)
\times\hat{p}^{\nu} - i\times m \times \hat{I}(x)] \times \hat{T}(x)
\times |> = 0,
\]
\begin{equation}
\{\hat{\gamma }^{\mu },\hat{}\hat{\gamma }^{\nu}\} =
\hat{\gamma }^{\mu}\times\hat{T}\times\hat{\gamma }^{\nu} +
\hat{\gamma }^{\nu}\times\hat{T}\times\hat{\gamma }^{\mu } =
2\times \hat{\eta}^{\mu\nu } \equiv 2\times g^{\mu\nu },
\hat{\gamma }^{\mu} = \hat{T}_{\mu\mu }^{1/2}\times
\gamma^{\mu}\times\hat{I}\;
({\rm no}\; {\rm sum}),
\label{eq:seven}\end{equation}

\vskip 0.5cm
Where $\gamma^{\mu  }$  are the  conventional  gammas and $\hat{\gamma
}^{\mu  }$ are    the  {\it isogamma   matrices}. Note  that   {\it the
anti-iso-commutators  of   the    isogamma matrices  yield   (twice)
the
Riemannian    metric g(x)},  thus   confirming  the  representation of
Einstein's (or other) gravitation in   the {\it structure} of  Dirac's
equation. As an illustration, we   have the {\it Dirac--Schwarzschild
equation}   given  by    Eq.s   (7)   with    $\hat{\gamma  }_{k}    =
(1-2M/r)^{-1/2}\times\gamma_{k}\times\hat{I}$ and  $\hat{\gamma }_{4}
=   (1-2M/r)^{1/2}\times\gamma^{4}\times\hat{I}$.   Similarly  one can
construct  the   isogravitational version of   all other  equations of
RQM.

These equations are not a mere mathematical curiosity because they
establish the compatibility of OIG with  experimental data in particle
physics in view   of the much  lower character  of gravitational  over
electromagnetic, weak and strong contributions. Our unification of the
special and   general   relativities is,  therefore,   compatible with
experimental evidence at both classical and operator levels.

The space components  $S\hat{O}(3)$, called {\it isorotations}$^{3i}$,
can be computed from isoexponentiations  (5) with the explicit form in
the  (x,y)--plane (were we  ignore again the factorization of $\hat{I}$
for simplicity)

\vskip 0.5cm
\[
x' =
x\times\cos(\hat{T}_{11}^{\frac{1}{2}}\times\hat{T}_{22}^{\frac{1}{2}}
\times\theta_{3}) -
y\times\hat{T}_{11}^{-\frac{1}{2}}\times\hat{T}_{22}^{\frac{1}{2}}
\times
\sin(\hat{T}_{11}^{\frac{1}{2}}\times\hat{T}_{22}^{\frac{1}{2}}
\times\theta_{3}),
\]
\begin{equation}
y' =
x\times\hat{T}_{11}^{\frac{1}{2}}\times\hat{T}_{22}^{-\frac{1}{2}}\times
\sin(\hat{T}_{11}^{\frac{1}{2}}\times\hat{T}_{22}^{\frac{1}{2}}\times
\theta_{3}) + y\times\cos(\hat{T}_{11}^{\frac{1}{2}}\times\hat{T}_{22}^
{\frac{1}{2}}\times\theta_{3}),
\label{eq:e8}\end{equation}

\vskip 0.5cm
(see$^{3s}$   for     general  isorotations    in    all there   Euler
angles).   Isotransforms  (8)   leave   invariant  all   ellipsoidical
deformations $x\times\hat{T}_{11}\times x + y\times\hat{T}_{22}\times y
+ z\times\hat{T}_{33}\times z = R$ of the sphere $x\times x + y\times y
+ z\times z = r$.
Such ellipsoid become perfect spheres $\hat{r}^{\hat{2}} =
(\hat{r}^{t} \times \hat{\delta } \times \hat{r}) \times \hat{I}_{s}$
in {\it isoeuclidean spaces}$^{3h,3r} \hat{E} (\hat{r}, \hat{\delta },
\hat{R}),
\hat{r} = \{\hat{r}^{k}\} = \{r^{k}\} \times \hat{I}_{s}, \hat{\delta }
=
\hat{T}_{s} \times \delta, \delta = Diag.(1,1,1), \hat{T}_{s} =
Diag.(\hat{T}_{11}, \hat{T}_{22}, \hat{T}_{33}), \hat{I}_{s} =
\hat{T}_{s}^{-1}$, called {\it isospheres}.

In fact, the deformation of the semi-axes $1_{k} \rightarrow
\hat{T}_{kk}$ while
the related units are deformed of the {\it inverse} amounts
$1_{k}\rightarrow\hat{T}_{kk}^{-1}$ preserves the perfect spheridicity
(because
the invariant in isospace is $(Length)^{2}\times (Unit)^{2})$. Note that
this
perfect sphericity in $\hat{E}$ is the geometric origin of the
isomorphism
$\hat{O}(3) \equiv O(3)$, with consequential preservation of the exact
{\it rotational} symmetry for the space--components $g(r)$ of all
possible
{\it Riemannian} metrics.

The connected isolorentz symmetry $S\hat{O}(3.1)$ is characterized by
the
isorotations and the {\it isolorentz boosts}$^{3h}$ which can be written
in the
$(3,4)$--plane

\vskip 0.5cm
\begin{eqnarray}
x^{3}{}'&=& x^{3}\times\sinh(\hat{T}_{33}^{\frac{1}{2}}\times
\hat{T}_{44}^{\frac{1}{2}}
\times v) - x^{4}\times\hat{T}_{33}^{-\frac{1}{2}}\times\hat{T}_{44}^
{\frac{1}{2}}\times
\cosh(\hat{T}_{33}^{\frac{1}{2}}\times\hat{T}_{44}\times       v)     =
\nonumber \\[0.5cm]
&& = \tilde{\gamma}\times
(x^{3}-\hat{T}_{33}^{-\frac{1}{2}}\times\hat{T}_{44}^{\frac{1}{2}}\times
\hat{\beta}\times  x^{4})
\nonumber \\[0.5cm]
x^{4}{}' &=& -x^{3}\times\hat{T}_{33}\times
c_{0}^{-1}\times\hat{T}_{44}^{-\frac{1}{2}}\times\sinh
(\hat{T}_{33}^{\frac{1}{2}}\times\hat{T}_{44}\times v)  +  x^{4}\times
\cosh (\hat{T}_{33}^{\frac{1}{2}}\times\hat{T}_{44}^{\frac{1}{2}}\times
v) = \nonumber \\[0.5cm]
&=&\tilde{\gamma }\times (x^{4}-\hat{T}_{33}^{\frac{1}{2}}
\times\hat{T}_{44}^{-\frac{1}{2}}\times\tilde{\beta}\times x^{3})
\nonumber \\[0.5cm]
\tilde{\beta} &=&
v_{k}\times\hat{T}_{44}^{\frac{1}{2}}/c_{0}\times\hat{T}_{44}^{\frac{1}{2}},\;\;\;
\tilde{\gamma} = (1-\tilde{\beta}^{2})^{-\frac{1}{2}}
.
\label{eq:e9}\end{eqnarray}

\vskip 0.5cm
Note that the above isotransforms are formally similar to the Lorentz
transforms,
as expected from their isotopic character. Isotransforms (9)
characterize the
{\it light isocone}$^{3s}$, i.e., the perfect cone in isospace
$\hat{M}$. In a
way similar to the isosphere, we have  the deformation of the light cone
axes
$1_{\mu} \rightarrow \hat{T}_{\mu\mu }$ while the corresponding units
are
deformed of the {\it inverse} amount $1_{\mu} \rightarrow
\hat{T}_{\mu\mu}^{-1}$.

In particular, the isolight cone also has the conventionals
characteristic
angle, as a
necessary condition for an isotopy (the proof of the latter property
requires
the use of isotrigonometric and isohyperbolic functions). Thus, {\it the
maximal
causal speed in isominkowski space is the conventional speed} $c_{0}$.
The
identity of the light cone and isocones is the geometric origin of the
isomorphism $S\hat{O}(3.1) \approx SO(3.1)$ and, thus, of the exact
validity of
the {\it Lorentz} symmetry for all possible {\it Riemannian} metrics
$g(x)$.

The {\it isotranslations} can be written
$x' = (\hat{e}^{i\times\hat{p}\times a})\hat{\times}\hat{x} =
[x + a\times A(x)]\times\hat{I},
\hat{p}' = (\hat{e}^{i\times\hat{p}\times a})\hat{\times}\hat{p} =
\hat{p}$,
where $
A_{\mu} = \hat{T}_{\mu\mu }^{1/2} + a^{\alpha }\times[\hat{T}_{\mu\mu
}^{1/2},\hat{}\hat{p}_{\alpha }]/1! + ....$
and they are also nonlinear, as expected.

Intriguingly, the isotopies identify
one additional symmetry which is absent in the conventional case. It is
here
called {\it isoselfscalar invariance} and it is given by the rescaling
of the unit
$\hat{I} \rightarrow\ \hat{I}' = n^{2}\times\hat{I}$, where n is
an 11-th parameter, under which the interval
remains invariant, $\hat{x}^{\hat{2}} =
(x^{\mu }\times\hat{T}_{\mu}^{\alpha }\times\eta_{\alpha\nu }\times
x^{\nu})\times\hat{I} \equiv
[x_{\mu}\times (n^{-2}\times\hat{T}_{\mu}^{\alpha})\times\eta_{\alpha\nu
}\times x^{\nu}]\times(n^{2}\times\hat{I})$.
Note that, even though $n^{2}$ is factorizable, the corresponding
isosymmetry
is not trivial, e.g., because $n^{2}$ enters into the {\it argument} of
the
isolorentz transforms $(9)$.

The same symmetry also holds for the isoinner
product (whenever $n$ does not depend on the integration variable),
$<\hat{\Phi}|\times\hat{T}\times |\hat{\Psi}>\times\hat{I} \equiv
<\hat{\Phi}|\times (n^{-2}\times\hat{T})\times
|\hat{\Psi}>\times(n^{2}\times{I})$.
Note finally that the latter symmetries have remained undetected
throughout this
century because they required the prior discovery of {\it new numbers},
those
with an arbitrary unit$^{3g}$.

\vskip 0.5cm

{\bf 6. Inclusion of interior gravitation.} The attentive
reader may have noted that the isotopies leave unrestricted the
functional dependence of the isometric. Its sole dependence on the
coordinates is therefore
a {\it restriction} which has been used so far for a representation of
{\it exterior
gravitation in vacuum}.

In the general case we have isometrics with an unrestricted
functional dependence,
$\hat{\eta} = \hat{T}(x,x,\partial \Psi, ...)\times\eta , \hat{T} > 0$,
which, as such, can represent {\it interior gravitation problems} with
an
{\it unrestricted nonlinearity in the velocities, wave functions and
their
derivatives}, as expected in realistic interior models, e.g., of neutron
stars,
quasars, black holes and all that.

Note also that the isometric can also contain
{\it nonlocal--integral terms}, e.g., representing
wave--overlappings$^{3s}$.
Nevertheless, the theory verifies the condition of locality in isospace,
called
{\it isolocality}, because its topology is everywhere local except at
the
unit$^{3t,4r}$.

The {\it general isopoincar\'{e} symmetry} is here defined as the
11--dimensional
set of {\it isorotations, isoboosts, isotranslations, isoinversions and
isoselfscalar transforms}. The {\it restricted isopoincar\'{e}
transforms} are
those in which the isounit is averaged into constants. The results of
this note
therefore imply the following:

\vskip 0.5cm
\newtheorem{teorema}{Theorem}
\begin{teorema}
.{\sl The 11-dimensional, general isopoincar\'{e} symmetry on
isominkowski spaces
over real isofields for well behaved and nowhere null isounits
constitutes the
largest possible isolinear, isolocal and isocanonical invariance
of isoseparation
$(2)$ for nonsingular isometrics with positive-definite isounits,
thus constituting the universal invariance of
exterior and interior gravitations}.
\end{teorema}

\vskip 0.3cm
The verification of the invariant under the isopoincar\'{e} transforms
of all
possible separation $(2)$ is instructive. The maximal character of the
isosymmetry can be proved as in the conventional case. Note that for any
arbitrarily given (diagonal) Riemannian metric $g(x)$ (such as
Schwarzschild,
Krasner, etc,) {\it there is nothing to compute} because one merely {\it
plots}
the $\hat{T}_{\mu\mu }$ terms in the decomposition
$g_{\mu\mu } = \hat{T}_{\mu\mu }\times\eta_{\mu\mu }$
(no sum) in the above given isotransforms. The invariance of the
separation
$x^{t}\times g\times x$
is then ensured. The $(2+2)$--de Sitter or other cases can be derived
from the
theorem via mere changes of signature or dimension of the isounit.

\vskip 0.5cm

{\bf 7. Concluding remarks.} In summary, in this note we have presented,
apparently for the first time, {\it a geometric unification of the
special and general relativities in both classical and operator
mechanics,
as well as for both exterior and interior problems.} The results
are centralluy dependent on the use of {\it isominkowskian geometry} as
introduced in this note,
rather than the use of the {\it isoriemannian} form as studied in
Ref.$^{3s}$.

The classical and operator geometric unification of
the special and general relativities for the exterior problem in vacuum
is
centrally depoendent on the achievement of a universal symmetry for
gravitation
which, by conception and construction, is locally isomorphic to the
Poincar\'{e}
symmetr of the special relativity. This eliminates the
historical difference between the special and general
relativities whereby the former admits a universal synnetry, while the
latter
does not$^{1,2}$. Note the {\it necessity} of the
representation of gravity in {\it isominkovski} space for the very
formulation
of its universal isopoincar\'{e} symmetry. In fact, no isosymmetry can
be constructed
in the Riemannian space, to our best knowledge.

The above occurrence has a number of implications. First, it
allow to illustrate the viewpoint expressed in Sect. 1 to
the effect that some of controversies in gravitation debated over
this century are not due to Einstein's field equations, but rather
to insufficiencies in the mathematics used for their treatment.

A typical case
is the controversy whether the total conservation laws of general
relativity
are compatible with those of the special relativity. Our representation
of
Einstein's equations via the novel isomathematics permits a resolution
of
this old controversy via a mere {\it visual examination}.

Recall that the generators of all space-time symmetries characterize
total
conserved quantities. The compatiobility of the total conservation laws
of the general and special relativities is therefore established by the
visual observation that {\it the generators of the Poincare' and
isopoincare'
symmetries coincide}. In fact, only the {\it mathematical operations} on
them
are changed in the transition from the relativistic to the gravitational
case.

The isominkowskian treatment of gravity also permits a resolution of
some of
the limitations of conventional gravitational models, such as their
insufficiency to provide an effective representation of {\it interior}
gravitational problems. In fact, conventional formulations of gravity
admit
only a limited dependence on the velocities, while being strictly
local-differential and derivable from a first-order Lagrangians
(variationally
self-adjoint$^{3d}$),
characteristics which are evidently exact for exterior propblems in
vacuum.

By comparison, interior gravitational problems, such as all forms of
gravitational collapse, are constitutred by extended and hyperdense
hadrons
in conditions of total mutual penetration in large numbers into small
regions of space. It is well known that these conditions imply effects
which
are {\it arbitrarily nonlinear in the
velocities as well as in the wavefunctions, nonlocal-integral on
various quantities and variationally nonselfadjoint}$^{3d,3e}$, (i.e.
not
representable via first-order Lagrangians). It is evident that the
latter
conditions are beyond any scientific expectation of quantitative
treatment
via conventional gravitational theories.

The isominkowskian formulation of gravity resolve this limitation too
and
shows that it is equally due to insufficiencies in the underlying
mathematics.
In fact, isogravitation extends the applicability of Einstein's axioms
to a form which is "direcly universal for
exterior and interior gravitations",
namely, capable of representing all exterior and interior conditions
considered (universality), directly in the x-frame of the experimenter
(direct universality), thus extending the above unification
to interior conditions.

As indicated earlier, this extension is due to the fact
that the funcional dependence of the metric in Riemannian treatments
is restricted
to the sole dependence on the local coordinates, g = g(x), while under
isotopies the same dependence becomes unrestricted,
$g = g(x, v, \phi, \partial {\psi}. ...)$ {\it without altering the
original
geometric axioms}. This results in {\it geometric unification of
exterior and interior problems}, despite their sizable structural
differences of
topological, analytic and other characters. The latter unification was
studied in details in ref.$^{3s}$  under
trhe {\it isoriemannian} geometry and it is studied with the
{\it isominkowskian} geometry in this note for the reasons indicated
earlier.

A first illustration of the extension of the axioms to
realistic interior conditions is offered by the isoselfscalar transforms
$\hat{T}_{\mu\mu } \rightarrow n^{-2}\times\hat{T}_{\mu\mu }$ which
permit the {\it representation of electromagnetic waves propagating
within
physical media with local varying speed} $c=c_{0}/n$. This allows the
construction,
apparently for the first time, of {\it Schwarzschild's and other
gravitational
models for the interior of atmospheres and chromospheres with a locally
varying
velocity of light}. Applications to specific cases, such as
gravitational
horizons, is then expected to permit refinements of current studies
evidently
due to deviations from the valud in vacuum of the speed of light in the
hyperdense chromospheres outside gravitational horizons.

Except for being well behaved (and non--null), the parameter n remains
unrestricted by the
isotopies and, therefore, $n=1$ in vacuum but otherwise it can be $n>1$
or $<1$.
As a result, {\it the isopoincar\'{e} symmetry is a natural invariance
for arbitrary
causal speeds, whether equal, smaller or bigger than the speed of
light}. The
former were known since Lorentz's$^{7a}$ times (see the related
quotation by
Pauli'$^{7b}$). The latter have been predicted since some time {\it in
interior
problems only}, but experimentally detected only recently, e.g., for the
speed
of photons traveling in certain guides$^{8a,8b}$ or for the speed of
matter in
astrophysical explosions$^{8c-8e}$. The recent Ref.$^{8f}$ has
identified
solutions of {\it conventional}
relativistic equations with {\it arbitrary speeds in vacuum} of
which $\hat{P}(3.1)$ is evidently the natural invariance).

Despite the local variation of $c$, the maximal causal speed on
$\hat{M}$ over
$\hat{R}$ remain $c_{0}$ again, because the change $c \rightarrow
c_{0}/n$ is
compensated by an inverse change of the unit. By recalling that the STR
is
evidently inapplicable (and not "violated") for arbitrary causal speeds,
we
can therefore say that {\it the isotopies render the STR universally
applicable,
not only for classical and operator gravitation, but also for arbitrary
causal
speeds}.

Also the light isocone remains applicable for interior gravitational
cases with arbitrary $c$. As such, the light isocone appears to be more
appropriate
than the conventional light cone for calculations, e.g., outside
gravitational
horizons which, being composed of hyperdense chromospheres, do not admit
the
conventional speed of light in vacuum $c_{0}$.

The indication of a number of developments currently under study appears
recommendable.
First, we note that the {\it the zeros of the space (time) component of
the isounit
represent gravitation horizons (singularities)}$^{3s}$. This
representation is
trivially equivalent to the conventional one for the {\it exterior} case
in vacuum.
However, gravitational collapse is a typical {\it interior} case for
which the
isotopic representation becomes nontrivial, e.g., because it permits the
inclusion
of the nonlinear, nonlocal and noncanonical effects indicated earlier.

Note that the zeros of the isounit have been excluded from Theorem 1
because of
their yet unknown topological structure.

Another aspect which is under study is the {\it
iso--grand--unification}$^{3s}$,
i.e., the inclusion of gravitation in the unified gauge theories of
electroweak
interactions via its embedding in the {\it unit} of the theory. If
successful,
these studies would extend the unification of this note to electroweak
interactions.

It should be also indicated that the isotopies with basic lifting
$I \rightarrow \hat{I}(x,\Psi ,...) = \hat{I^{\dagger}}$
constitute only the first step of a chain of generalized methods$^{3f}$.
The
second class is given by the {\it genotopies}$^{3a}$ in which the
isounit is no
longer Hermitean. This broader class geometries in a natural way the
interior
irreversibility and it has been used, e.g., for the black hole model of
ref.$^{4d}$. The third class of methods is given by the (multi--valued)
{\it hyperstructures}, in which the generalized unit is constituted by a
{\it set of non--Hermitean quantities}. The latter most general known
class
appears to be particularly significant for quantitative studies of
biological
structures in which the conventional RQM is manifestly inapplicable due
to its
reversibility.

Also, the isotopies, genotopies and hyperstructures admit
antiautomorphic
images, called {\it isodualities}, and charactrized by the map
$\hat{I} \rightarrow -\hat{I}^{d} = -\hat{I}^{\dagger}$ which are
currently
under study for antimatter$^{3q}$. In this case the energy--momentum
tensor of
antimatter becomes {\it negative--definite}, thus removing a problem of
compatibility between the current representations of antimatter in
classical and
particle physics. The gravitational treatment of antimatter via the
isodualities of the isominkowskian geometry will be studied elsewhere.

On historical grounds, we note that, as studied in detail in
memoir$^{3t}$
for the general case of RHM, our OIG can be interpreted as a
{\it nonunitary completion} of RQM considerably along the historical
$E-P-R$
argument$^{9a}$ for which yon Neumann theorem$^{9b}$ and Bell's
inequalities$^{9c}$
do not apply evidently because of its nonunitary structure.

Moreover, from the
abstract identity of the right modular associative action
$H\times |\Psi >$ and its isotopic image $\hat{H}\hat{\times }|\hat{\Psi
}>$, one can see that the
isoeigenvalue equation
$\hat{H}\hat{\times }|\hat{\Psi }> = E_{\hat{T}}\times |\hat{\Psi}>$
characterizes an explicit and concrete {\it operator realization of the
"hidden
variable"} $\lambda = \lambda (x,...)\equiv \hat{T}$. Our isotopic
formulation
of gravity can therefore be interpreted as a realization of the theory
of hidden
variables. After all, the "hidden" character of gravitation in our
theory is
illustrated by the recovering of the conventional unit under the
isoexpectation
value $\hat{<}\hat{I}\hat{>} = I$.

In conclusion, the viewpoint we have attempted to convey in this note is
that
an axiomatically consistent operator formulation of gravity {\it always}
existed.
It did creep in un--noticed until now because embedded where nobody
looked for,
in the {\it unit} of RQM.

\vskip 0.5cm

{\bf Acknowledgments.} The autor would like to
thank for invaluable comments the participants
to: the {\sl 7 Marcel Grossmann Meeting  on General Relativity} held at
Stanford university in July   1994; the {\sl International  Workshops}
held at the Istituto   per la Ricerca di  Base  in Molise,   Italy, on
August  1995 and  May 1996;  and the  {\sl  International Workshop  on
Physical Interpretation of Relativity  Theories} held at the  Imperial
College, London, on September 1996. Moreover,
the author would like to express his sincere appreciation
to the Editors of
{\it Rendiconti Circolo Matematico Palermo},
{\it Foundations of Physics} and {\it Mathematical Methods in Applied
Sciences},
for invaluable, penetrating and
constructive critical comments in the editorial processing of
memoirs in the field, without which this paper could not have seen
the light.

\end{document}